\documentclass[aps,amsmath,twocolumn,amssymb,floatfix,showpacs,superscriptaddress,nofootinbib,longbibliography]{revtex4-1}
\usepackage{MnSymbol}
\usepackage{braket}
\usepackage[dvipsnames]{xcolor}
\usepackage{float}
\usepackage{subfigure}
\usepackage{tikz}
\usepackage[colorlinks=true,linktoc=page,linkcolor=OliveGreen,citecolor=purple,urlcolor=violet]{hyperref}

\mathchardef\mhyphen="2D 

\newcommand{\ie}{{i.e.,\,\,}}
\newcommand{\eg}{{e.g.,~}}

\newcommand\bea{\begin{eqnarray}}
\newcommand\eea{\end{eqnarray}}
\newcommand\beq{\begin{equation}}  
\newcommand\eeq{\end{equation}}

\newcommand{\non}{\nonumber}  
\usepackage[normalem]{ulem}
\definecolor{lime}{HTML}{A6CE39}
\usepackage{sidecap,tikz}
\DeclareRobustCommand{\orcidicon}{\hspace{-1.0mm}
	\begin{tikzpicture}
		\draw[lime, fill=lime] (0.0,0.0) 
		circle [radius=0.15] 
		node[white] {{\fontfamily{qag}\selectfont \tiny \,ID}};
		\draw[white, fill=white] (-0.0525,0.095) 
		circle [radius=0.007];
	\end{tikzpicture}
	\hspace{-3.0mm}
}
\foreach \x in {A, ..., Z}{\expandafter\xdef\csname orcid\x\endcsname{\noexpand\href{https://orcid.org/\csname orcidauthor\x\endcsname}{\noexpand\orcidicon}}
}

\AtBeginDocument{%
	\newwrite\bibnotes
	\def\bibnotesext{Notes.bib}
	\immediate\openout\bibnotes=\jobname\bibnotesext
	\immediate\write\bibnotes{@CONTROL{REVTEX41Control}}
	\immediate\write\bibnotes{@CONTROL{%
			apsrev41Control,author="08",editor="1",pages="1",title="1",year="1"}}
	\if@filesw
	\immediate\write\@auxout{\string\citation{apsrev41Control}}%
	\fi
}%

\begin{document}

\title{Spin-orbit coupling driven topological superconductivity in twisted bilayer graphene-WSe$_2$ heterostructures}  

\author{Kamalesh Bera\orcidA{}}
\email{kamalesh.bera@iopb.res.in}
\affiliation{Institute of Physics, Sachivalaya Marg, Bhubaneswar-751005, India}
\affiliation{Homi Bhabha National Institute, Training School Complex, Anushakti Nagar, Mumbai 400094, India}

\author{Arijit Saha\orcidC{}}
\email{arijit@iopb.res.in}
\affiliation{Institute of Physics, Sachivalaya Marg, Bhubaneswar-751005, India}
\affiliation{Homi Bhabha National Institute, Training School Complex, Anushakti Nagar, Mumbai 400094, India}

\author{Tanay Nag\orcidB{}}
\email{tanay.nag@hyderabad.bits-pilani.ac.in}
\affiliation{Department of Physics, BITS Pilani-Hyderabad Campus, Telangana 500078, India}

\begin{abstract}
Commencing from the low-energy Bistritzer–MacDonald continuum model of twisted bi-layer graphene (tBLG) with proximity-induced Ising, Rashba, and intrinsic spin-orbit couplings (SOCs), we construct the corresponding Bogoliubov-de Gennes Hamiltonian with conventional $s$-wave pairing and theoretically investigate 
the emergence of topological superconductivity in it. The latter can possibly be experimentally demonstrated in tBLG, Niobium and tungsten diselenide heterostructures. The topological superconducting phases, bearing an effective $p$-wave pairing profile and exhibiting inverted band dispersion, are protected by a bulk gap and are topologically characterized by Chern numbers. In the absence of intrinsic SOC, variation of the twist angle and other SOC strengths yield extended gapless, trivial, and topological phases, with phase boundaries exactly matching the closing of direct band gaps. The topological regime exhibits clear band inversion in the combined particle-hole and spin space, along with distinct Bloch localization profiles compared to the trivial phase. Including intrinsic SOC generates additional topological phases and eliminates the gapless phase, indicating the enhanced stability of the gapped topological superconducting regime
in tBLG.

\end{abstract}

\maketitle

\textcolor{blue}{\textit{Introduction:}-} Moir\'e systems have emerged as a highly tunable material platform in which the twist angle acts as a new control parameter, enabling the formation of nearly dispersionless flat bands at magic angle from the otherwise highly dispersive graphene bands~\cite{Santos-Peres-tBLG,Shallcross-tBLG,MacDonald-tBLG,Koshino-tBLG}. Consequently, near the magic angle in twisted bilayer graphene (tBLG), a variety of strongly correlated phases, including unconventional superconductivity, correlated insulating states, magnetism, and Chern insulating phases etc.~\cite{Cao2018-unconv_sc,Cao2018-corr_insulator,Oh2021-tBLG(xpt),Nuckolls2020-tBLG(xpt),Magnetism-tBLG,CIS-tBLG,MI+SC-tBLG,Wu2021-chernIns-expt,
All_Magic_Angle-tBLG,Origin_of_Magic_Angle-tBLG,Heavy_fermion-tBLG} are observed. Extending beyond tBLG, a broad family of moir\'e materials have been realized through different layer numbers, stacking arrangements, lattice structures, and twist configurations, encompassing twisted multilayer graphene systems and transition metal dichalcogenide (TMD) moir\'e hybrid structures~\cite{Mono-Bi-graphene1,Park2021-tTLG1,Liu2020-tDBLG1,tDBLG2,He2021-tDBLG3,KB_transport_tBG,Adak2022-tDBLG6,Chakraborty_2022,Sinha2022,KB_tDBLG,
TMD-homobilayers1,KB_NH_hBN_tBG,rTTLG+hBN1,bera2025chiral,tBG_alpha_t3}.

On the other hand, topological superconductivity hosting Majorana fermions has emerged as a distinct quantum phase of matter that supports zero-energy quasiparticles such as Majorana fermions, which are anticipated to play a central role in future technologies, particularly realizing fault-tolerant topological quantum computation~\cite{TI_TSC_RMP_review_Zhang,Sato_2017,RMP_Nayak,TSC_classification_Schnyder,TSC_review_Kallin_2016,kitaev2009periodic}. Among the various proposed routes, one of the most extensively studied platforms for realizing Majorana zero modes is a semiconducting nanowire with strong spin-orbit coupling (SOC) subjected to a Zeeman field and placed in close proximity to a conventional $s$-wave superconductor~\cite{RNW_thr1,RNW_thr2,RNW_thr3,RNW_thr4,PhysRevB.108.L081403,PhysRevB.107.035427}. Although significant experimental progress has been made, and several studies have reported indirect signatures consistent with Majorana bound states (MBSs)~\cite{RNW_expt1,RNW_expt2}, definitive evidence providing unambiguous confirmation remains elusive~\cite{RNW_expt3}. This ongoing challenge has motivated broad theoretical and experimental efforts to identify alternative platforms for topological superconductivity (TSC). These include magnetic impurity chains~\cite{HSC_thr1,HSC_thr2,HSC_thr3,Imp_chn1,Imp_chn2,Imp_chn_expt1,Imp_chn_expt2}, two-dimensional semiconducting heterostructures~\cite{2D_SOC_Bz1,2D_SOC_Bz2,2D_SOC_Bz3,2D_SOC_Bz_expt1,2D_SOC_Bz_expt2}, planar Josephson junctions~\cite{TSC_JJ1,TSC_JJ2,TSC_JJ_expt}, graphene-based systems~\cite{TSC_graphene1,TSC_graphene2,TSC_graphene3,TSC_graphene4}, TMD-based systems~\cite{TSC_TMD1,TSC_TMD2,TSC_TMD3}, and several other emerging architectures.

In literature, TMDs (\eg MoS$_2$, WSe$_2$ etc.) in proximity to graphene are known to significantly enhance the SOC strength in it~\cite{WSe2_Graphene1, WSe2_Graphene2, WSe2_Graphene3, WSe2_Graphene4, WSe2_Graphene5, WSe2_Graphene6}. Apart from Rashba SOC, another type of SOC, known as Ising SOC, also emerges. This acts as an effective Zeeman field with opposite signs in the two valleys~\cite{ISOC1,ISOC2,ISOC3}. In tBLG as well, several theoretical and experimental studies have demonstrated the induction of topology, ferromagnetism, and enhanced stability of superconductivity~\cite{WSe2_tBLG_topology, WSe2_tBLG_ferro, WSe2_tBLG_SC1, WSe2_tBLG_SC2} due to the proximity induced SOCs. While intrinsic TSC in tBLG and other moir\'e systems has been extensively investigated~\cite{DMFT_TSC_tBLG,RG_TSC_tBLG,RG_TSC_tDBLG,MFT_TSC_tBLG,QMC_TSC_tBLG,BdG_TSC_tBLG}, there exist only a few proposals concerning TSC in proximity-coupled heterostructures in tBLG. In some earlier studies, TSC in tBLG was realized within an atomistic model by considering ferromagnet/tBLG/superconductor heterostructures~\cite{TSC_tBLG_Lado1,TSC_tBLG_Lado2,TSC_tBLG_Lado3}. Here, following the low-energy continuum model, we propose an alternative route to engineer TSC in tBLG by utilizing the Ising SOC induced via WSe$_2$ as an example. In other words, our quest is to engineer an effective $p$-wave pairing starting from a conventional $s$-wave superconductor in the tBLG-WSe$_2$ heterostructures.

In this article, we theoretically investigate an experimentally possible realization of TSC in tBLG-WSe$_2$ heterostructure placed in proximity to a regular $s$-wave superconductor. Starting from the low-energy continuum model with Ising and Rashba SOCs induced by WSe$_2$, we construct the corresponding Bogoliubov-de Gennes (BdG) Hamiltonian for tBLG with $s$-wave superconducting pairing in it. The topological superconducting phases, exhibiting band inversion, are protected by a bulk gap and characterized by the Chern number. Therefore, this signals towards the emergent $p$-wave nature of the superconducting gap. In the absence of intrinsic SOC, the topological phase diagram obtained by tuning the twist angle and Ising SOC reveals extended gapless phases at larger twist angles and trivially gapped phases at smaller angles, while increasing Rashba SOC drives a transition from trivial to topological superconducting phases. Notably, the closing of the direct band gap precisely coincides with the phase boundaries separating distinct topological sectors. The topological regime further exhibits a pronounced band inversion in the combined particle-hole and spin space of the bulk BdG spectrum, accompanied by distinctly different Bloch localization profiles compared to the trivial phase. Upon incorporating intrinsic SOC, the phase landscape becomes significantly richer with the emergence of additional topological phases absent in the previous case, while the gapless regime entirely disappears, indicating enhanced stability of the gapped topological superconducting spectrum. 

\begin{figure}[t]
	\centering
	\subfigure{\includegraphics[width=0.48\textwidth]{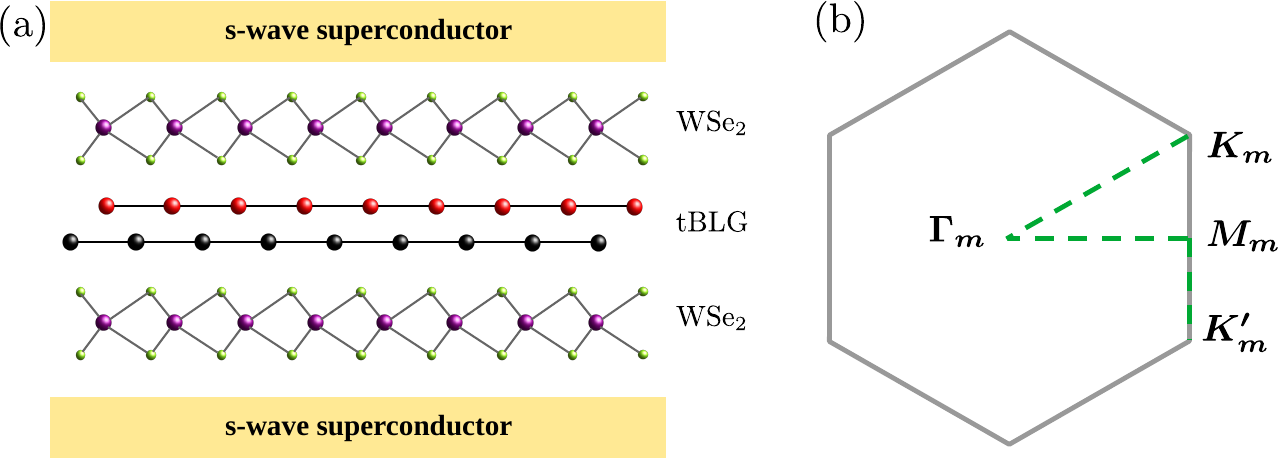}}
	\caption{(a) A simplified schematic representation of our setup is presented. The side views of tBLG with black and red color lines representing two layers of graphene involved in realizing tBLG. The latter is encapsulated within WSe$_{2}$ from both sides. Finally, the combined system is placed between two regular $s$-wave superconductors. In panel (b), we represent the hexagonal moir\'e Brillouin zone (mBZ) and the high symmetry path along which the band structure is shown (see latter text for discussion).
	}
	\label{schematic}
\end{figure}

\textcolor{blue}{\textit{Model:}-} In this section, we present the low-energy continuum model used to investigate the emergence of topological superconductivity (TSC) in tBLG. To realize the proposed topological phase, we consider a tBLG$-$$\mathrm{WSe}_{2}$ heterostructure (schematically shown in Fig.~\ref{schematic}) placed in close proximity to a conventional $s$-wave superconductor.
To begin with, we introduce the low-energy continuum Hamiltonian for such a heterostructure~\cite{MacDonald-tBLG,Koshino-tBLG,WSe2_tBLG_topology}. In the basis $\Phi_{K_{\xi},\bf{q}} = (A_{q_1 \uparrow}, B_{q_1 \uparrow}, A_{q_1 \downarrow}, B_{q_1 \downarrow}, A_{q_2 \uparrow}, B_{q_2 \uparrow}, A_{q_2 \downarrow}, B_{q_2 \downarrow},)^{T}$
which incorporates both sublattice and spin degrees of freedom, the Hamiltonian can be written as
\begin{eqnarray}
	H^{\xi}_{0} (\mathbf{q})=\left(\begin{array}{cc}%
		h^{\xi}({\mathbf{q_1}}) & T^{\xi}\\
		{T^{\xi}}^\dagger & h^{\xi}({\mathbf{q_2}})
	\end{array}\right)\ ,
	\label{Eq:tBLG}
\end{eqnarray}
here, $h^{\xi}(\mathbf{q}_1)$ and $h^{\xi}(\mathbf{q}_2)$ represent the low-energy Hamiltonians for the two graphene monolayers forming tBLG, with $\mathbf{q}_{l} = R[(-1)^{l}(\theta/2)]\mathbf{q}$, with $\mathbf{q}=(q_x,q_y)$,  denoting the momentum in layer $l$ near the valley $K_{\xi}$. The term $T^{\xi}$ describes the interlayer coupling between the two layers. Further details of the continuum model, along with the corresponding parameter values used in our calculations, are provided in the supplemental material (SM) \cite{supp}.
The low-energy Hamiltonian for a single graphene layer in proximity to a $\mathrm{WSe}_{2}$ substrate, inducing SOC near the valley $K_{\xi}$, can be expressed as~\cite{TSC_graphene4,WSe2_tBLG_topology}
\begin{eqnarray} 
	h^{\xi}({ \bf q}) &=& \hbar v_{F} \big( \xi q_{x} \sigma_{x} - q_{y} \sigma_{y} \big) + \lambda_{R} \big( \xi  s_{x} \sigma_{y} - s_{y} \sigma_{x}\big)\non  \\
	&& + \lambda_{I} s_{z} \sigma_{0} + \xi \lambda_{K} s_{z} \sigma_{z}\ .
	\label{Eq:SLG_with_SOC}
\end{eqnarray}
Here, the Pauli matrices $\sigma$ and $s$ act on the sublattice and spin degrees of freedom, respectively. The first term represents the low-energy Dirac Hamiltonian of monolayer graphene, while the remaining terms capture the proximity-induced SOCs. Specifically, $\lambda_{R}$, $\lambda_{I}$, and $\lambda_{K}$ denote the strengths of Rashba, Ising, and intrinsic SOCs~\cite{Kane-mele_QSH1,Kane-mele_QSH2}, respectively. Note that, under the low-energy approximation in the honeycomb lattice, the SOC terms are momentum independent (see supplementary materials (SM) for a detailed derivation of the low-energy effective 
Hamiltonian~\cite{supp}). Although, Rashba SOC induces momentum-dependent spin splitting of the flat bands in tBLG, it alone cannot break the protecting $C_{2}\mathcal{T}$ symmetry of the gapless Dirac points and therefore cannot open up a topological gap. In contrast, Ising SOC acts as an effective valley-dependent Zeeman field, producing opposite spin polarization in opposite valleys and breaking $C_{2}\mathcal{T}$ symmetry within each valley~\cite{WSe2_tBLG2}. This opens up a possible topological gap that is further stabilized by the intrinsic SOC. Nevertheless, the full system, including both valleys ($K_{\xi}=K$ and $K^{\prime}$), remains invariant under spinful time-reversal symmetry (TRS), $\mathcal{T}=i\xi_x s_y \sigma_0 \mathcal{K}$, where $\xi_x$ is the $x$-component of the Pauli matrix acting in valley space and $\mathcal{K}$ 
denotes complex conjugation.

As the $K$ and $K^{\prime}$ valleys remain connected via TRS, this enables the BdG Hamiltonian to be formulated in a basis where the particle and hole sectors are composed of states originating from opposite valleys. 
Hence, we write the BdG Hamiltonian in the following Nambu basis $\Psi_{\bf{q}} = \bigl( \Phi_{K, \bf{q}\uparrow}, \Phi_{K, \bf{q}\downarrow},\Phi^{\dagger}_{K^{\prime},-\bf{q}\uparrow}, \Phi^{\dagger}_{K^{\prime}, -\bf{q}\downarrow} \bigr)^{T}$ as 
$H = \frac{1}{2} \sum_{\bf{q}} \Psi_{\bf{q}}^{\dagger} H_{\text{BdG}}(\mathbf{q}) \Psi_{\bf{q}}$, with
\begin{align}
	H_{\text{BdG}} (\mathbf{q}) = 
	\left( \begin{array}{cc}
		H^{\xi}_{0} (\mathbf{q}) & H_{\Delta}\\
		H_{\Delta}^\dagger & -\mathcal{T} H^{\xi}_{0} (\mathbf{q}) \mathcal{T}^{-1} 
	\end{array}\right)\ .\
	\label{tBLG_BdG}
\end{align}
where, $H^{\xi}_{0} (\mathbf{q})$ represents the low-energy continuum Hamiltonian of the tBLG-$\mathrm{WSe}_{2}$ heterostructure near valley-$K_{\xi}$. 
Also, $H_{\Delta}$ describes the superconducting pairing potential. We further note that the explicit form of the pairing potential is given by
\begin{align}
	H_{\Delta}=\left(\begin{array}{cc}%
		h^{(1)}_{\Delta} & 0\\
		0 & h^{(2)}_{\Delta}
	\end{array}\right)\ .
	\label{Eq:H:AB-AB-Full}
\end{align}	 

Here, the $h^{(l)}_{\Delta}$ denotes the proximity induced conventional $s$-wave superconducting pairing in each layer-$l$ of tBLG and are given by $h^{(l)}_{\Delta} =  i s_{y} \sigma_{0} \Delta_{\text{sc}}$, where, $\Delta_{\mathrm{sc}}$ represents the superconducting order parameter.
Since the interlayer coupling, arising from van der Waals interactions, is much weaker than the characteristic intralayer energy scale~\cite{tBLG-2016,tBLG-Gate-voltage}, we focus exclusively on superconducting pairing within individual layers. This approximation is motivated by the clear distinction of energy scales and is expected to capture the dominant contribution to the emergent topological superconducting state.


\begin{figure}[t]
	\centering
	\subfigure{\includegraphics[width=0.49\textwidth]{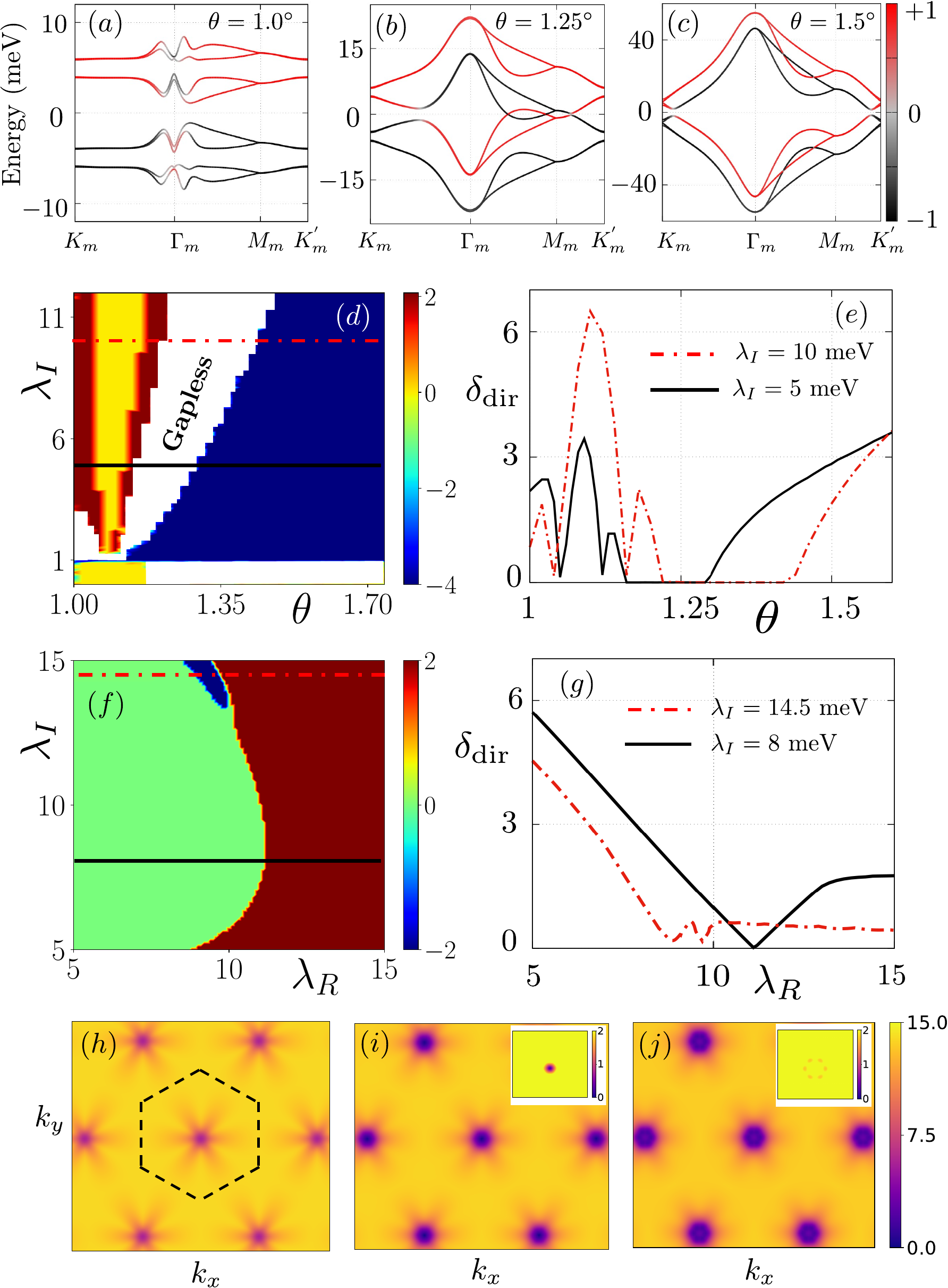}}
	\caption{The bulk band structure obtained from the BdG Hamiltonian of tBLG is demonstrated along the high symmetry path in panels (a), (b), (c), corresponding to the twist angles $\theta = 1.0^{\circ}, 1.25^{\circ}, 1.5^{\circ}$, respectively, for fixed values of Ising SOC $\lambda_{I} = 5$ meV, Rashba SOC $\lambda_{R} = 10$ meV. The colorbar represents $\langle \tau_{z}s_{z}\rangle$ for each of the moir\'e bands.  In panel (d), we depict the topological phase diagram with Chern number ($\mathcal{C}$), represented as colorbar in $\theta$ - $\lambda_{I}$ plane for constant values of $\lambda_{R} = 10$ meV. In panel (e), we display the direct band gap along the $\lambda_{I} = 5$ meV (solid black line) and $\lambda_{I} = 10$ meV (dashed red line) lines of $\theta$ - $\lambda_{I}$ phase diagram. In panel (f), we represent another topological phase diagram, representing $\mathcal{C}$ as the colorbar, at the magic angle in $\lambda_{R}$ - $\lambda_{I}$ plane. The corresponding direct band gap closing transitions are shown in panel (g) with respect to $\lambda_{R}$. In panels (h), (i), and (j), we showcase the density plot for band gap (in meV) on the reciprocal space, respectively for $\lambda_{R} = 5, 11.1$ and $15$ meV choosing the twist angle $\theta = 1.05^{\circ}$ and $\lambda_{I} = 8$ meV. For all the above panels, we set the superconducting order parameter to unity \ie $\Delta_{\mathrm{sc}} = 1$ meV.} 
	\label{Ising_Rashba_theta_TSC}
\end{figure}

\textcolor{blue}{\textit{Topological superconductivity mediated by Ising SOC:}-} We here present and analyze our results on TSC in small-angle tBLG proximitized by WSe$_2$ and a conventional $s$-wave superconductor. To illustrate the emergence of the TSC phase, we first examine the evolution of the band dispersion obtained from the BdG Hamiltonian (\ie Eq.~(\ref{tBLG_BdG})) for fixed finite values of Ising SOC ($\lambda_{I}$), Rashba SOC ($\lambda_{R}$), and superconducting pairing at different twist angles. In Figs.~\ref{Ising_Rashba_theta_TSC}(a,b,c), we depict the corresponding band dispersions for $\theta = 1.0^{\circ}, 1.25^{\circ},$ and $1.5^{\circ}$, respectively, along the high-symmetry path ($K_{m}-\Gamma_{m}-M_{m}-K_{m}^{'}$ in the mBZ). Clear signatures of band inversion are visible in the combined particle-hole and spin weights, captured by $\langle \tau_{z}s_{z}\rangle$. Although, both $\theta = 1.0^{\circ}$ and $1.5^{\circ}$ exhibit band inversion, the corresponding inversion occurs near $\Gamma_m$ for the former and near $K_m$ and $K_m^{'}$ for the latter, leading to distinct 
TSC phases. As shown later in the phase diagram, the larger twist angle exhibits a doubling of Chern number compared to the smaller one. In contrast, the band dispersion for $\theta = 1.25^{\circ}$ corresponds to a gapless phase.

Having established the band inversion, we then investigate the topological character of the superconducting state via the Chern number $\mathcal{C}$ associated with the BdG Hamiltonian described in Eq.~(\ref{tBLG_BdG}). In Fig.~\ref{Ising_Rashba_theta_TSC}(d), we present the topological phase diagram in the $(\theta-\lambda_I)$ parameter space for fixed Rashba SOC strength $\lambda_R=10$ meV and superconducting pairing amplitude $\Delta_{\text{sc}}=1$ meV. 
The Chern number is evaluated using a gauge-invariant formalism that computes the total Chern number below charge neutrality (see SM~\cite{supp} for details). The resulting phase diagram reveals that varying the twist angle alone can drive multiple topological superconducting transitions even at fixed $\lambda_{I}$, 
reflecting the strong twist-angle dependence of the underlying moiré band structure. Since changing $\theta$ modifies the bandwidth and relative band separation, it alters the interplay between superconducting pairing and SOC-induced band reconstruction, causing repeated bulk gap closings and reopenings that generate distinct topological phases with different Chern numbers as shown in Fig.~\ref{Ising_Rashba_theta_TSC}(e). Near the magic angle ($\theta \approx 1.05^\circ$), 
we find a trivial superconducting phase with $\mathcal{C}=0$ (yellow region), surrounded by a $\mathcal{C}=2$ TSC phase (red region). At larger twist angles, 
an extended gapless region appears before entering into another TSC phase with $\mathcal{C}=-4$ (blue region). This higher-Chern phase indicates gap closings 
at both the $K$ and $K_m$ points with opposite chirality compared to the $\mathcal{C}=2$ phase, where only $\Gamma_m$ participates in band inversion. 
The intermediate gapless regime hosts nodal loops within the mBZ together with a nontrivial momentum-space spin texture (see SM~\cite{supp} for details).

Note that, the above mentioned phase transitions are supported by the analysis of the direct band gap which is defined as the minimum energy difference between the first conduction band and the first valence band over the entire mBZ.  To be precise, the direct band gap is given by $\delta_{\rm dir} = \min_{\mathbf k\in \rm mBZ} \bigl[\,\epsilon_{1}(\mathbf k) - \epsilon_{-1}(\mathbf k)\bigr]$ where, $\epsilon_{-1}(\mathbf k)$ and $\epsilon_{1}(\mathbf k)$ denote the energies of the highest valence band and the lowest conduction band, respectively, at momentum $\mathbf{k}$. In Fig.~\ref{Ising_Rashba_theta_TSC}(e),
we find that the direct band gap becomes vanishingly small for $\theta \approx 1.05^{\circ}$, and $1.12^{\circ}$ ($\theta \approx 1.04^{\circ}$, 
and $1.16^{\circ}$) when $\lambda_{I} = 5$ meV, 
($\lambda_{I} = 10$ meV), 
identifying the topological phase transitions from $\mathcal{C} = 2 \rightarrow 0$ and $\mathcal{C} = 0 \rightarrow 2$. On the other hand, the extended gapless region manifests the vanishing nature of the direct band gap over a substantial window of $\theta$ that increases with enhancing $\lambda_{I}$ and matches well with the phase diagram displayed in Fig.~\ref{Ising_Rashba_theta_TSC}(d). After that, the gapped region corresponds to TSC phase with $\mathcal{C} = -4$.

To exhibit the tunability of the topological phases with Rashba SOCs, we present the corresponding phase diagram in Fig.~\ref{Ising_Rashba_theta_TSC}(f) 
in the $\lambda_{R}$ - $\lambda_{I}$ plane at magic angle $\theta = 1.05^{\circ}$ with $\Delta_{\text{sc}} = 1$ meV. While most of the region with $\lambda_{R} < 10$ meV remains topologically trivial (green region) which is consistent with the trivial region depicted in  Fig.~\ref{Ising_Rashba_theta_TSC}(d). For $\lambda_{R} > 10$ meV, the system enters TSC phases characterized by  $\mathcal{C} = 2$ (red region) and $\mathcal{C} = -2$ (blue region). Notably, the $\mathcal{C} = 2$ phase persists over an extended region of the phase diagram. To identify the topological phase transitions, we calculate the direct band gap 
as a function of $\lambda_{R}$ along the lines $\lambda_{I} = 8$ meV (solid black line) and $\lambda_{I} = 14.5$ meV (dashed red line), as shown in Fig.~\ref{Ising_Rashba_theta_TSC}(g). To further clarify the nature of the topological transition, we display the gap-closing behavior over the mBZ by calculating the energy difference between the first conduction and the first valence bands throughout the mBZ. In Figs.~\ref{Ising_Rashba_theta_TSC}(h,i,j), 
we display the band gap distribution across the mBZ for $\lambda_{R}=5$, $11.1$, and $15$ meV, respectively, at fixed $\lambda_{I}=8$ meV (solid black line) 
near the magic angle. We find that the gap-closing transition occurs at $\lambda_{R}=11.1$ meV, where the gap 
closes only at the $\Gamma_m$ point, indicating that topological phase transitions at lower twist angles are primarily driven by the $\Gamma_m$ point 
(see Fig.~\ref{Ising_Rashba_theta_TSC}(i)). Although, both Figs.~\ref{Ising_Rashba_theta_TSC}(i,j) appear nearly gapless, the insets (zoomed-in view near 
the gap minimum) clearly exhibit that the actual gap closing transition occurs only in Fig.~\ref{Ising_Rashba_theta_TSC}(i).

\begin{figure}
	\centering
	\subfigure{\includegraphics[width=0.49\textwidth]{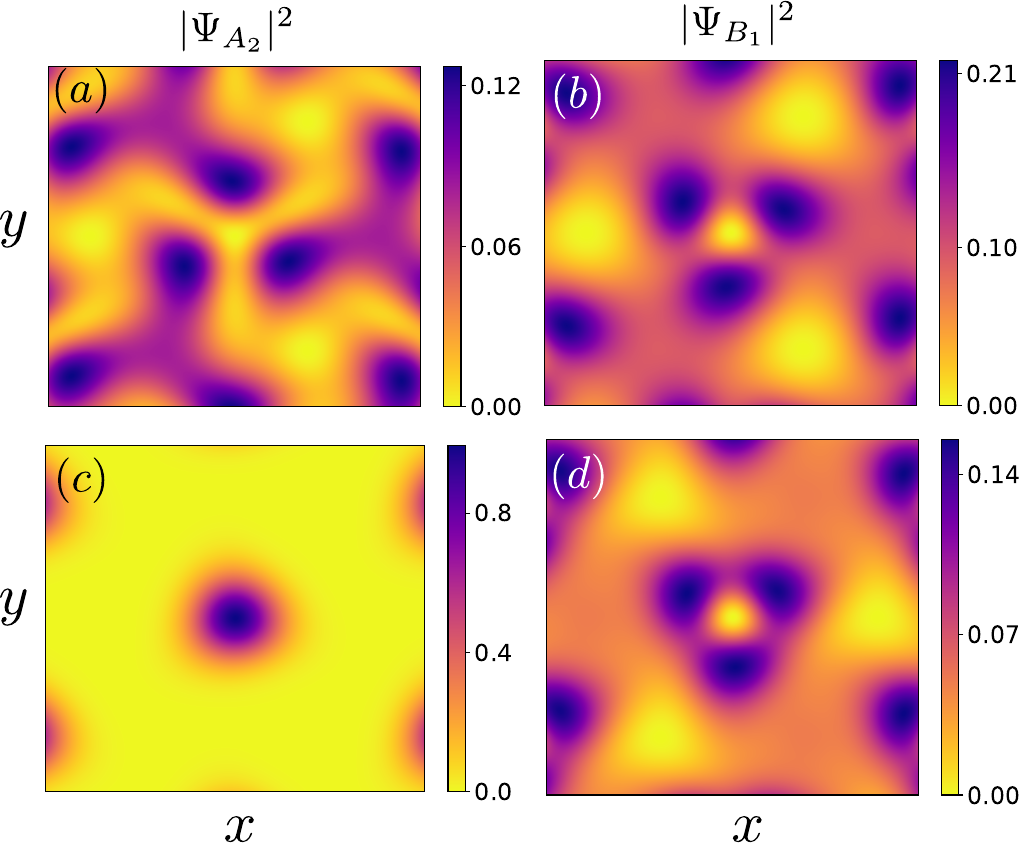}}
	\caption{Density plot for the squared amplitude of Bloch states are illustrated in the $x$-$y$ plane for tBLG at a twist angle $\theta = 1.05^{o}$ corresponding to the $\Gamma_{m}$-point of the first valence band. The two columns refer to the sublattice-A of layer-2 ($|\psi_{A_{2}}|^{2}$) and sublattice-B of layer-1 ($|\psi_{B_{1}}|^{2}$). The first row (a,b) and second row (c,d) correspond to $\lambda_{R} = 5$ meV \ie topologically trivial and $\lambda_{R} = 15$ meV \ie topologically non-trivial regions, respectively. We consider fixed Ising SOC $\lambda_{I} = 8$ meV and superconducting order parameter $\Delta_{\mathrm{sc}} = 1$ meV. 	
	}
	\label{tBLG_Bloch_states}
\end{figure}

Finally, we investigate the real-space spatial profiles of the modulo squared Bloch-state amplitudes, shown in Figs.~\ref{tBLG_Bloch_states}(a,c), and (b,d) 
for sublattice A on the second layer and sublattice B on the first layer, respectively. We focuss on the flat conduction band near the $\Gamma_{m}$ point, 
(see SM~\cite{supp} for more details). This analysis provides insight into the evolution of the electronic states across different topological phases. Note that, Figs.~\ref{tBLG_Bloch_states}(a,b) [(c,d)]correspond to $\lambda_{R} = 5$ [$\lambda_{R} = 15$] meV, representing the trivial $\mathcal{C} = 0$ [$\mathcal{C} = 2$] phase as depicted by green [brown] region in Fig.~\ref{Ising_Rashba_theta_TSC}(f).  As evident from the Bloch localization, distinct topological phases exhibit markedly different Bloch-state localization profiles, with the trivial phase displaying comparatively weaker localization than the topological phase. 
Interestingly, $|\Psi_{B_1}|^2$ remains qualitatively unaltered irrespective of the band topology of the model, while $|\Psi_{A_2}|^2$ changes its profile significantly between the topological and trivial phases. Therefore, it is not generally true that Bloch state amplitude profiles can always differentiate between topological and trivial phases~\cite{bera2025chiral}. Consequently, one should not rely solely on such local probes; instead, a proper topological characterization requires the detailed analysis of an appropriate invariant, such as the Chern number which we have already analysed  before.

\begin{figure}[t]
	\centering
	\subfigure{\includegraphics[width=0.49\textwidth]{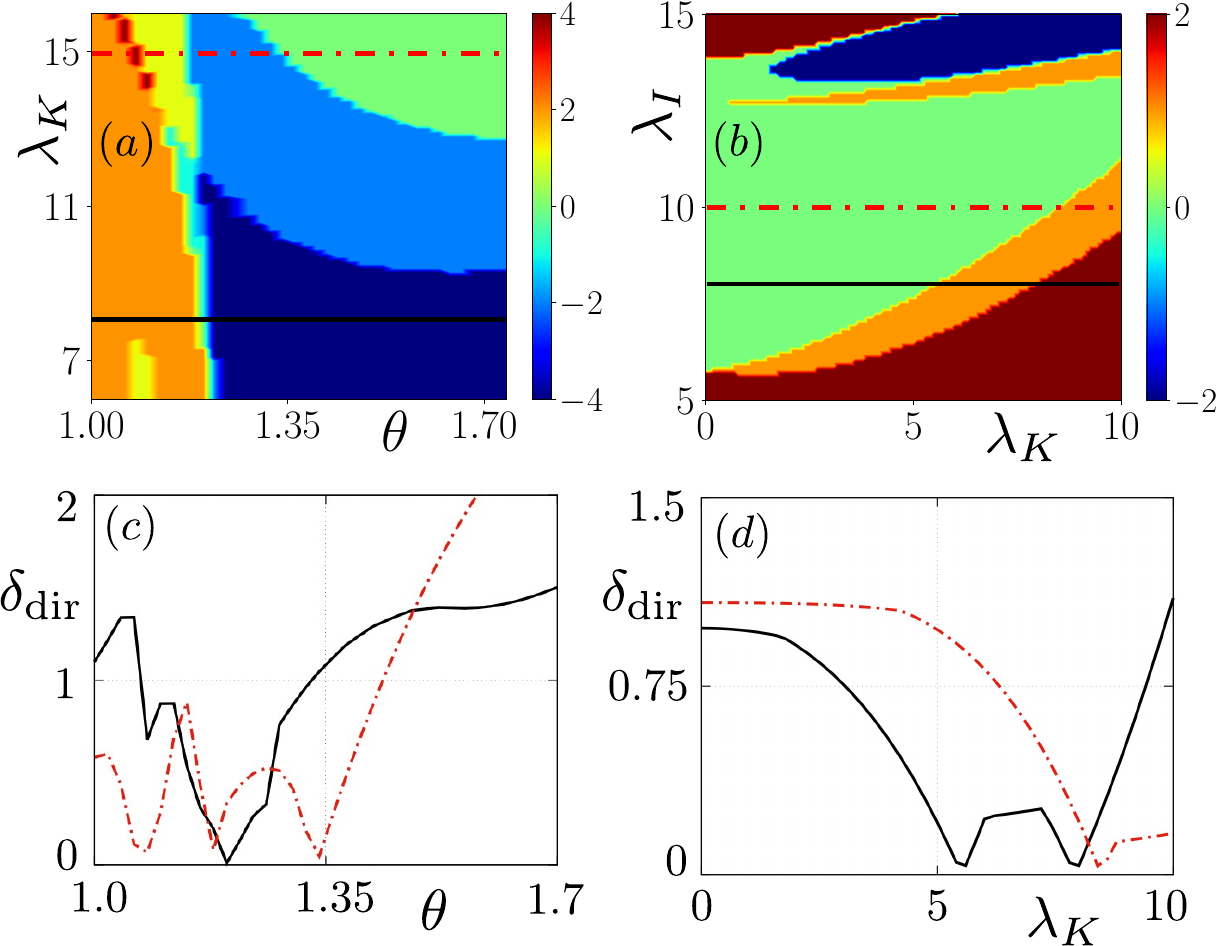}}
	\caption{The topological phase diagram, representing Chern number $\mathcal{C}$, is demonstrated in panel (a) in the $\theta$-$\lambda_{K}$ plane with fixed value of Ising SOC $\lambda_{I} = 5$ meV and Rashba SOC $\lambda_{R} = 10$ meV. In panel (b), we depict the phase diagram in the $\lambda_{K}$-$\lambda_{I}$ plane with $\theta = 1.05^{o}$ and $\lambda_{R} = 10$ meV. Direct band gap \ie $\delta_{\text{dir}}$ in meV scale is displayed as a function of the twist angle $\theta$, and intrinsic SOC $\lambda_{K}$, respectively, in  panels (c) and (d) choosing two distinct choices of $\lambda_K$ and $\lambda_I$ 
	(indicated by solid and dashed lines in panels (a) and (b)).  We consider superconducting order parameter $\Delta_{\mathrm{sc}} = 1$ meV.
	}
	\label{tBLG_intrinsic_phase}
\end{figure}

\textcolor{blue}{\textit{Effect of Intrinsic SOC:}-} Here, we discuss the impact of intrinsic SOC on the emerging TSC phase of tBLG. Although intrinsic SOC alone cannot induce a topological superconducting phase unlike monolayer graphene~\cite{Kane-mele_QSH1,Kane-mele_QSH2}, it can significantly modify the existing phases originating from the simultaneous presence of Ising and Rashba SOCs. Hence, similar to earlier discussion, we explore the following phase diagrams: (i) in the $\lambda_K$-$\theta$  plane for fixed Ising and Rashba SOC strengths  
(see Fig.~\ref{tBLG_intrinsic_phase}(a)); and (ii) in the $\lambda_I$-$\lambda_K$ plane at magic angle, with fixed Rashba SOC strengths and superconducting order parameter (see Fig.~\ref{tBLG_intrinsic_phase}(b)).

Investigating Fig.~\ref{tBLG_intrinsic_phase}(a), we find that at the higher twist-angle regime ($\theta > 1.20^{\circ}$), three extended gapped phases emerge.  Two topologically non-trivial phases with $\mathcal{C}=-4$ (deep blue) and $\mathcal{C}=-2$ (light blue), along with one topologically trivial phase with $\mathcal{C}=0$ (green). In contrast, for smaller twist angles, the dominant topological superconducting phases are characterized by $\mathcal{C}=2$ (orange) and $\mathcal{C}=1$ (yellow). To further analyze these phases and the associated topological phase transitions, we calculate the direct band gap with respect to twist angle $\theta$ along two representative reference lines highlighted in Fig.~\ref{tBLG_intrinsic_phase}(a): $\lambda_{K}=15$ meV (red dashed line) and $\lambda_{K}=8$ meV (solid black line). As shown in Fig.~\ref{tBLG_intrinsic_phase}(c), the direct band gap along the red dashed line approaches zero three times, corresponding to successive topological transitions from $\mathcal{C}=2 \rightarrow 1$, $\mathcal{C}=1 \rightarrow -2$, and $\mathcal{C}=-2 \rightarrow 0$. In contrast, the solid black line exhibits only one single topological phase transition, corresponding to $\mathcal{C}=2 \rightarrow -4$.

At the magic angle, we demonstrate the phase diagram in $\lambda_I$-$\lambda_K$ plane as shown in Fig.~\ref{tBLG_intrinsic_phase}(b), where we observe that an extended region of the parameter space remains non-topological (green region). However, there also exists regions with non-trivial band topology characterized by $\mathcal{C}=1$ (orange region), $\mathcal{C}=2$ (red region), and $\mathcal{C}=-2$ (deep blue region). Referring along $\lambda_{I}=10$ meV (dashed red line) and $\lambda_{I}=8$ meV (solid black line) in the phase diagram of Fig.~\ref{tBLG_intrinsic_phase}(b), we showcase the direct band gap as a function of intrinsic SOC $\lambda_{K}$ in Fig.~\ref{tBLG_intrinsic_phase}(d). We observe that the $\lambda_{I}=8$ meV line crosses two topological phases, and the two corresponding phase transitions ($\mathcal{C}=0 \rightarrow 1$ and $\mathcal{C}=1 \rightarrow 2$) are consistently indicated by two points in the direct band gap behavior where it touches zero (see Fig.~\ref{tBLG_intrinsic_phase}(d)). A similar analysis holds for the $\lambda_{I}=10$ meV line along which we have one topological phase transition associated with $\mathcal{C}=0 \rightarrow 1$.

In contrast to the earlier discussion, where intrinsic SOC is absent, we find that the inclusion of that type of SOC not only drastically modifies the existing phase boundaries but also introduces new TSC phases. Specifically, the $\mathcal{C}=1$ phase was not observed in the phase diagram explored previously in the absence of $\lambda_K$ for moderate value of twist angle (see Fig.~\ref{Ising_Rashba_theta_TSC}(f)). 
Furthermore, gapless region appears to be forbidden in presence of intrinsic SOC. Therefore, intrinsic SOC provides more robust gap protection that, in turn, assists in stabilizing the topological nature of the model. It is important to note that the TSC is caused by an underlying $p$-wave nature of the effective superconducting order (see SM~\cite{supp} for more details). The topological phase transitions are accompanied by the band inversion, where the combined particle-hole and spin weights, captured by $\langle \tau_{z}s_{z}\rangle$, are exchanged between valence and conduction moir\'e bands. However, in principle, band inversions can also occur in terms of the weights $\langle s_{z}\sigma_{z}\rangle$ and $\langle \tau_{z}\sigma_{z}\rangle$, in addition to $\langle \tau_{z}s_{z}\rangle$. 



\textcolor{blue}{\textit{Summary and Conclusions:}-} To summarize, in this article, we demonstrate the realization of TSC in a tBLG-WSe$_2$ heterostructure placed in close proximity to a conventional $s$-wave superconductor. Starting from the low-energy Bistritzer-MacDonald type of continuum model for tBLG, incorporating proximity-induced Ising, Rashba and intrinsic SOCs from WSe$_2$, we construct the corresponding BdG Hamiltonian with conventional $s$-wave pairing. Using this framework, we systematically analyze the emergence of TSC through band dispersion, direct band gap, and Chern number calculations. We first chart out topological phase diagrams in the absence of intrinsic SOC by varying the twist angle with Ising SOC  where extended gapless and trivially gapped phases are obtained for higher and lower twist angles, respectively. On the other hand, trivial and topological phases appear for weak and strong Rashba SOC. We find exact correspondence between the vanishing of direct band gaps and boundaries between the two distinct phases. Interestingly, the bulk BdG band spectrum in the topological regime manifests clear band inversion in the combined particle-hole and spin space. We also find a markedly different profile of Bloch localization amplitudes between trivial and topological phases, which can be considered as a secondary indicator only. We extend the above analysis in the presence of intrinsic SOC, where a variety of additional topological phases appear that are otherwise absent in the previous case. Importantly, gapless phase disappears 
in the presence of intrinsic SOC, indicating the stability of the gapped TSC spectrum. 



As far as experimental connection of our proposal is concerned, we note that proximity-induced Ising and Rashba SOCs can attain strengths upto $\sim 20$ meV~\cite{WSe2_tBLG,WSe2_Graphene4,WSe2_tBLG2}, comparable to the narrow bandwidth of tBLG~\cite{MacDonald-tBLG}. This allows the experimental viability of our work.  The large lattice mismatch between graphene and WSe$_2$ leads only to a weak short-period moiré modulation~\cite{WSe2_tBLG_SC1}, whose effect is expected to be negligible and is therefore ignored in our analysis. Our results thus may be possible to experimentally tested in the near future considering heterostructures consisting of tLBG, $s$-wave superconductors and TMD based systems.

\textcolor{blue}{\textit{Acknowledgments:}-} 
K.B. and A.S. acknowledge SAMKHYA: High-Performance Computing Facility provided by Institute of Physics, Bhubaneswar, and the two workstations provided by the Institute of Physics, Bhubaneswar from the DAE APEX project for numerical computations. T.N. and A.S. thank the Advanced Research Grant (ARG) from Anusandhan National Research Foundation Grant No. ANRF/ARG/2025/003163/PS.

\textcolor{blue}{\textit{Data Availability Statement:}-}The datasets generated and analyzed during the current study are available from the authors
upon reasonable request.


\bibliography{bibfile.bib}{}

\normalsize\clearpage
\begin{onecolumngrid}
\begin{center}
		{\fontsize{12}{12}\selectfont
			\textbf{Supplemental Material for ``Spin-orbit coupling driven topological superconductivity in twisted bilayer graphene-WSe$_2$ heterostructures''\\[5mm]}}
		{\normalsize  Kamalesh Bera\orcidA{},$^{1,2,*}$ Arijit Saha\orcidC{},$^{1,2,*}$  and  Tanay Nag\orcidB{},$^{3}$ \\[1mm]}
		{\small $^1$\textit{Institute of Physics, Sachivalaya Marg, Bhubaneswar-751005, India}\\[0.5mm]}
		{\small $^2$\textit{Homi Bhabha National Institute, Training School Complex, Anushakti Nagar, Mumbai 400094, India}\\[0.5mm]}
		{\small $^3$\textit{Department of Physics, BITS Pilani-Hyderabad Campus, Telangana 500078, India}\\[0.5mm]}
\end{center}
	
\newcounter{defcounter}
\setcounter{defcounter}{0}
\setcounter{equation}{0}
\renewcommand{\theequation}{S\arabic{equation}}
\setcounter{figure}{0}
\renewcommand{\thefigure}{S\arabic{figure}}
\setcounter{page}{1}
\pagenumbering{roman}

\renewcommand{\thesection}{S\arabic{section}}

\section{DERIVATION OF EFFECTIVE LOW-ENERGY HAMILTONIAN FOR GRAPHENE WITH SPIN-ORBIT COUPLINGS}\label{Sec:I}
In this section, we discuss how various spin-orbit coupling (SOC) terms appear in the low-energy continuum description of a single layer of graphene. Previous works~\cite{graphene_SOCs1, graphene_SOCs2} have extensively studied the effects of SOC in graphene starting from the real-space tight-binding Hamiltonians. 
Here, for completeness, we briefly outline the derivation of the low-energy effective Hamiltonian used in our model construction, beginning instead from the 
momentum-space tight-binding formulation. We consider the Hamiltonian in the basis $(A_{k\uparrow}, B_{k\uparrow}, A_{k\downarrow}, B_{k\downarrow})^{T}$, which 
can be written as
\begin{align}
	H({\bf k}) = 
	\left(\begin{array}{cccc}
		\lambda_{A} f_{I}(\mathbf{k}) & -t f(\mathbf{k}) & 0 & -i \lambda_{R} f_{R}({\bf k})\\
		-t f(\mathbf{k})^{*} & -\lambda_{B} f_{I}(\mathbf{k}) & i \lambda_{R} f_{R}({-\bf k}) & 0\\ 
		0 & -i \lambda_{R} (f_{R}({-\bf k}))^{*} & -\lambda_{A} f_{I}(\mathbf{k}) & -t f(\mathbf{k})\\
		i\lambda_{R} (f_{R}({\bf k}))^{*} & 0 & -t f(\mathbf{k})^{*} & \lambda_{B} f_{I}(\mathbf{k}) \\ 
	\end{array}\right)\ ,
	\label{Eq:S1}
\end{align}	
Here, $t$ denotes the nearest-neighbor hopping amplitude, while $\lambda_{R}$ represents the Rashba SOC strength associated with nearest-neighbor spin-flipping hopping processes. The latter leads to spin splitting in the band structure. The parameters $\lambda_{A}$ and $\lambda_{B}$ correspond to the sublattice-resolved intrinsic SOC strengths induced by the proximity effect due to WSe$_{2}$. In the following, we discuss how these SOC terms give rise to intrinsic (Kane-Mele)~\cite{Kane-mele_QSH1, Kane-mele_QSH2} and Ising-type SOCs. The associated functions in Eq.~(\ref{Eq:S1}) are given by~\cite{graphene_SOCs1, graphene_SOCs2},
\begin{align*}
	f({\bf k}) &= 1 + 2 e^{\frac{3 i k_{y}}{2}} \cos \left(\frac{\sqrt{3} k_{x} }{2}\right)\ , \\
	f_{R}({\bf k}) &= \frac{2}{3} \left(1 - e^{\frac{3 i k_{y}}{2}} \Bigg[\cos\Bigg(\frac{\sqrt{3} k_{x}}{2}\Bigg) - \sin\Bigg(\frac{\sqrt{3} k_{x}}{2}\Bigg) \Bigg]\right)\ , \\
	f_{I}({\bf k}) &= -\frac{2}{3 \sqrt{3}} \left(\sin (\sqrt{3} k_{x})-2 \sin \Bigg(\frac{\sqrt{3} k_{x}}{2}\Bigg) \cos \Bigg(\frac{3 k_{y}}{2}\Bigg)\right)\ . 
\end{align*}
		
The Dirac points of single-layer graphene are located at $K = \left(\frac{4\pi}{3\sqrt{3}}, 0\right)$ and $K^{'} = \left(-\frac{4\pi}{3\sqrt{3}},0\right)$ 
in the Brillouin zone (BZ). To obtain the low-energy continuum description of graphene in the presence of such SOCs, we perform a Taylor expansion around the valley-$K$ point as
\begin{align*}
	f ({\bf k})\Big|_{K+q} \simeq -\frac{3 }{2} (q_{x} + i q_{y}), \hspace{20pt}
	f_{R}({\bf -k})\Big|_{K+q} &\simeq   2 , \hspace{20pt}  f_{R}({\bf k})\Big|_{K+q} \simeq   (q_{x} - i q_{y}), \hspace{10pt} \text{and} \hspace{10pt}
 	f_{I}({\bf k})\Big|_{K+q} \simeq 1\ .
\end{align*}
Retaining only the leading-order terms and defining the Fermi velocity as $v_{F}=3t/2$ (assuming the carbon-carbon bond length $a=1$ and $\hbar=1$), one can obtain
\begin{align}
	h({\bf q}) = 
	\left( \begin{array}{cccc}
		\lambda_{A} & v_{F} q_{+} & 0 & 0\\
		v_{F} q_{-} & -\lambda_{B} & 2 i \lambda_{R} & 0\\ 
		0 & -2 i \lambda_{R} & -\lambda_{A} & v_{F} q_{+}\\
		0 & 0 & v_{F} q_{-} & \lambda_{B} \\ 
	\end{array}\right)\ ,
	\label{Eq:S2}
\end{align}	
From the above Hamiltonian, we observe that when $\lambda_{A}=\lambda_{B}$, the corresponding SOC term reduces to the intrinsic SOC ($\lambda_{K}$)~\cite{Kane-mele_QSH1,Kane-mele_QSH2}, whose finite strength lifts the degeracy at the Dirac points resulting in a mass gap. In contrast, when $\lambda_{A}=-\lambda_{B}$, the spin degeneracy is lifted, producing an effective Zeeman-like field commonly referred to as valley-Zeeman or Ising SOC ($\lambda_{I}$). Therefore, for the more general case where $\lambda_{A}\neq\lambda_{B}$, these SOCs can be conveniently rewritten as $\lambda_{K}=(\lambda_{A}+\lambda_{B})/2$ and $\lambda_{I}=(\lambda_{A}-\lambda_{B})/2$. Equivalently, one can express $\lambda_{A}=\lambda_{K}+\lambda_{I}$ and $\lambda_{B}=\lambda_{K}-\lambda_{I}$, following which the Hamiltonian in Eq.~(\ref{Eq:S2}) can be recast in the form~\cite{WSe2_Graphene3},
\begin{align}
	h({\bf q}) &= 
	\left(\begin{array}{cccc}
		\lambda_{K} + \lambda_{I} & v_{F} q_{+} & 0 & 0\\
		v_{F} q_{-} & -\lambda_{K} + \lambda_{I} & 2 i \lambda_{R} & 0\\ 
		0 & -2 i \lambda_{R} & -\lambda_{K} - \lambda_{I} & v_{F} q_{+}\\
		0 & 0 & v_{F} q_{-} & \lambda_{K} - \lambda_{I} \\ 
	\end{array}\right)\ , \\
	&= v_{F} \big(q_{x} \sigma_{x} - q_{y} \sigma_{y} \big) + \lambda_{R} \big( s_{x} \sigma_{y} - s_{y} \sigma_{x}\big)
   + \lambda_{K} s_{z} \sigma_{z} + \lambda_{I} s_{z} \sigma_{0}\ .
	\label{Eq:S3}
\end{align}	
This corresponds to the compact form of the single-layer graphene Hamiltonian with various SOCs used throughout the main text. Here, we have explicitly shown 
the derivation only around the valley-$K$; the corresponding calculation for valley-$K^{\prime}$ follows the same procedure as mentioned above. 

\section{Bistritzer-MacDonald model in presence of SOCs}\label{Sec:II}
Twisted bilayer graphene (tBLG) is formed by stacking two monolayer graphene sheets with a relative rotational misalignment between them. In this section, we briefly describe the construction of our model Hamiltonian based on the Bistritzer-MacDonald framework~\cite{MacDonald-tBLG,Koshino-tBLG} which we have used in our analysis. 

The primitive lattice vectors of monolayer graphene are defined as $\mathbf{a}_{1} = a(3/2,\sqrt{3}/2)$ and $\mathbf{a}_{2} = a(3/2,-\sqrt{3}/2)$, where $a = 0.142~\mathrm{nm}$ denotes the carbon-carbon distance in graphene. The corresponding reciprocal lattice vectors are given by $\mathbf{b}_{1} = (2\pi/a)(1/3,1/\sqrt{3})$ and $\mathbf{b}_{2} = (2\pi/a)(1/3,-1/\sqrt{3})$. To construct the twisted bilayer system, we rotate the two graphene layers relative to each other such that layer-1 and layer-2 are rotated by angles $-\theta/2$ and $\theta/2$, respectively. Here, the layer index $l=1,2$ labels the two graphene sheets. Under this rotation, the reciprocal lattice vectors transform as $\mathbf{b}^{(l)}_{1,2} = R[(-1)^{l} (\theta / 2)] \mathbf{b}_{1,2}$ where $R(\phi)$ represents the two-dimensional rotation matrix corresponding to an angle $\phi$. For small twist angles, the slight mismatch between the two rotated lattices generates a long-wavelength moir\'e superlattice. The associated moir\'e reciprocal lattice vectors can be obtained from the momentum difference between the two layers and are expressed as $\mathbf{G}^{m}_{1,2} = \mathbf{b}^{(1)}_{1,2} - \mathbf{b}^{(2)}_{1,2}$. After rotation, the Dirac points of the individual graphene layers are shifted in momentum space and adopt the form $K^{(l)}_{\xi} = -\xi (2 \mathbf{b}^{(l)}_{1} + \mathbf{b}^{(l)}_{2})/3$, where $\xi=\pm1$ denotes the valley index corresponding to the $K$ and $K^{\prime}$ valleys, respectively.

Here, we introduce the low-energy continuum Hamiltonian for tBLG in the presence of proximity-induced spin--orbit couplings (SOCs) originating from the adjacent WSe$_2$ layers. The Hamiltonian is expressed in the basis $\Phi_{K_{\xi},\bf{q}} = (A_{q_1 \uparrow}, B_{q_1 \uparrow}, A_{q_1 \downarrow}, B_{q_1 \downarrow}, A_{q_2 \uparrow}, B_{q_2 \uparrow}, A_{q_2 \downarrow}, B_{q_2 \downarrow})^{T}$ which incorporates both the sublattice and spin degrees of freedom for the two graphene layers. The corresponding continuum Hamiltonian is given by
\begin{eqnarray}
	H_{0}^{\xi}({\mathbf{q}}) =\left(\begin{array}{cc}%
		h^{\xi}({\mathbf{q_1}}) & T^{\xi}\\
		{T^{\xi}}^\dagger & h^{\xi}({\mathbf{q_2}})
	\end{array}\right)\ ,
	\label{A1}
\end{eqnarray}
where, $h^{\xi}({\mathbf{q_1}})$ and $h^{\xi}({\mathbf{q_2}})$ represent the effective Hamiltonians for the two rotated monolayer graphene sheets near the valley-$K_{\xi}$. Here, the momenta are defined as $\mathbf{q}_{(l)} = R[(-1)^{(l)} (\theta/2)]\mathbf{q}$. The term $T^{\xi}$ describes the interlayer tunneling between the two graphene layers.
	
The low-energy effective Hamiltonian for a single graphene layer in proximity to a WS$e_{2}$ substrate, in the vicinity of the valley-$K_{\xi}$ point, 
is given by
\begin{eqnarray} 
	h^{\xi}({ \bf q}) &=& \hbar v_{F} \big( \xi q_{x} \sigma_{x} - q_{y} \sigma_{y} \big) + \lambda_{R} \big( \xi  s_{x} \sigma_{y} - s_{y} \sigma_{x}\big) + \lambda_{I} s_{z} \sigma_{0} + \xi \lambda_{K} s_{z} \sigma_{z}\ ,
	\label{A2}
\end{eqnarray}
In this Hamiltonian (Eq.~(\ref{A2})), the Pauli matrices $\sigma$ and $s$ operate in the sublattice and spin spaces, respectively. The first term represents the low-energy Dirac Hamiltonian of the single layer graphene, whereas the other terms account for the spin-orbit interactions induced via the proximity effect. 
In particular, the parameters $\lambda_{R}$, $\lambda_{I}$, and $\lambda_{K}$ characterize the strengths of the Rashba, Ising, and intrinsic (Kane-Mele type)~\cite{Kane-mele_QSH1,Kane-mele_QSH2} SOCs, respectively.
	
The interlayer coupling between the two layers (\ie tBLG) is given by the matrix $T$ as follows~\cite{MacDonald-tBLG, Koshino-tDBLG},
\begin{eqnarray}
	T^{\xi} = T_{1,\xi} + T_{2,\xi} e^{i\xi \mathbf{G_{1}^{m}} \cdot \mathbf{r}} + T_{3,\xi} e^{i\xi (\mathbf{G_{1}^{m}} + \mathbf{G_{2}^{m}}) 
		\cdot \mathbf{r}}\ ,
	\label{A3}
\end{eqnarray}
where, $\mathbf{G^{m}} = n_{1}\mathbf{G_{1}^{m}} + n_{2}\mathbf{G_{2}^{m}}$ is the reciprocal lattice vector in the  moiré Brillouin zone (mBZ) (the superscript $m$ in $\mathbf{G}$'s denotes that) and $n_{1}$, $n_{2}$ are integers. The tunneling matrices in Eq.~(\ref{A3}) are given by
\begin{eqnarray}
	T_{i,\xi} &= s_{0} \otimes t_{i,\xi}
\end{eqnarray}
with, 
\begin{eqnarray}
	t_{1,\xi} &=
	\left( \begin{array}{cc}
		u  & u^{\prime} \\
		u^{\prime} & u
	\end{array}\right) ~~,~~
	t_{2,\xi} 
	&=
	\left( \begin{array}{cc}
		u  & u^{\prime} \omega^{-\xi} \\
		u^{\prime}\omega^{\xi} & u
	\end{array}\right) ~~,~~ 
	t_{3,\xi} 
	=
	\left( \begin{array}{cc}
		u  & u^{\prime} \omega^{\xi} \\
		u^{\prime}\omega^{-\xi} & u
	\end{array}\right)\ .
	\label{A4}
\end{eqnarray}
In our calculations, we use $\hbar v_{F}/a = 2135.4$ meV~\cite{RMP-Graphene-review,McCann_2013-BilayerReview,Koshino-tBLG}. The parameters $u$ and $u^{\prime}$ represent the interlayer tunneling amplitudes between the AA/BB and AB/BA sublattices, respectively. To incorporate the effects of out-of-plane lattice corrugation, we choose $u = 79.7$ meV and $u^{\prime} = 97.5$ meV~\cite{Koshino-tBLG,corrugation-DFT-uu,Dai2016-uu}. Here, $\omega = e^{2 \pi i/3}$. Furthermore, in the numerical calculations for the tBLG, we impose a momentum-space cutoff of $4|\mathbf{G}^{m}_{1,2}|$ within the momentum lattice of the mBZ.

\section{Realization of effective $p + ip$-pairing from $s$-wave superconductivity}\label{Sec:III}
In this section, we investigate the nature of the effective pairing responsible for the appearance of topological superconducting phase in tBLG-WSe$_2$ heterostructures. However, obtaining an analytical treatment of the Bistritzer-MacDonald model, even at the lowest-order approximation, becomes extremely challenging in the presence of SOCs and superconductiing pairing. Therefore, to gain analytical insight, we focus instead on the individual graphene layers constituting the tBLG system.

The Bogoliubov-de Gennes (BdG) Hamiltonian for a single layer of graphene in presence of SOCs and superconducting gap can be written in the basis $\bigl( A_{q\uparrow}, B_{q\uparrow}, A_{q\downarrow}, B_{q\downarrow}, A^{\dagger}_{-q\uparrow}, B^{\dagger}_{-q\uparrow}, A^{\dagger}_{-q\downarrow}, B^{\dagger}_{-q\downarrow} \bigr)^{T}$ as,
\begin{align}\label{SLG_BdG}
H_{BdG} =	\left(
	\begin{array}{cccccccc}
		\lambda_{I} & v_{F} q_{+} & 0 & 0 & 0 & 0 & \Delta_{\text{sc}}  & 0 \\
		v_{F} q_{-} & \lambda_{I} & 2 i \lambda_{R} & 0 & 0 & 0 & 0 & \Delta_{\text{sc}}  \\
		0 & -2 i \lambda_{R} & -\lambda_{I} & v_{F} q_{+} & -\Delta_{\text{sc}}  & 0 & 0 & 0 \\
		0 & 0 & v_{F} q_{-} & -\lambda_{I} & 0 & -\Delta_{\text{sc}}  & 0 & 0 \\
		0 & 0 & -\Delta_{\text{sc}}  & 0 & -\lambda_{I} & -v_{F} q_{+} & 0 & 2 i \lambda_{R} \\
		0 & 0 & 0 & -\Delta_{\text{sc}}  & -v_{F} q_{-} & -\lambda_{I} & 0 & 0 \\
		\Delta_{\text{sc}}  & 0 & 0 & 0 & 0 & 0 & \lambda_{I} & -v_{F} q_{+} \\
		0 & \Delta_{\text{sc}}  & 0 & 0 & -2 i \lambda_{R} & 0 & -v_{F} q_{-} & \lambda_{I}\  \\
	\end{array}
	\right)\ ,
\end{align}
In Eq.~(\ref{SLG_BdG}), all terms except $\Delta_{\text{sc}}$, which denotes the conventional $s$-wave pairing strength, have already been discussed in Sec.~\ref{Sec:I}. We now introduce a unitary transformation $U$ to obtain a dual Hamiltonian corresponding to the BdG Hamiltonian (\ie Eq.~(\ref{SLG_BdG})) as
\begin{eqnarray}
	U = \frac{1}{\sqrt{2}}\left(
	\begin{array}{cc}
		I_{4} & I_{4} \\
		-I_{4} & I_{4} \\	
	\end{array}
	\right)\ ,
\end{eqnarray}
where, $I_{4}$ is a 4-by-4 identity matrix.

Therefore, upon performing the dual transformation, the Hamiltonian can be recast into the following dual form as,
\begin{eqnarray}
	H_{BdG}^{D} = U^{\dagger} H_{BdG} U\ ,
\end{eqnarray}
where the dual Hamiltonian is given by,
\begin{align}\label{SLG_BdGD}
H_{BdG}^{D} = \left(
\begin{array}{cccccccc}
	0 & 0 & 0 & i \lambda_{R} & \lambda_{I} & v_{F} q_{+} & \Delta_{\text{sc}}  & -i \lambda_{R} \\
	0 & 0 & i \lambda_{R} & 0 & v_{F} q_{-} & \lambda_{I} & i \lambda_{R} & \Delta_{\text{sc}}  \\
	0 & -i \lambda_{R} & 0 & 0 & -\Delta_{\text{sc}}  & -i \lambda_{R} & -\lambda_{I} & v_{F} q_{+} \\
	-i \lambda_{R} & 0 & 0 & 0 & i \lambda_{R} & -\Delta_{\text{sc}}  & v_{F} q_{-} & -\lambda_{I} \\
	\lambda_{I} & v_{F} q_{+} & -\Delta_{\text{sc}}  & -i \lambda_{R} & 0 & 0 & 0 & i \lambda_{R} \\
	v_{F} q_{-} & \lambda_{I} & i \lambda_{R} & -\Delta_{\text{sc}}  & 0 & 0 & i \lambda_{R} & 0 \\
	\Delta_{\text{sc}}  & -i \lambda_{R} & -\lambda_{I} & v_{F} q_{+} & 0 & -i \lambda_{R} & 0 & 0 \\
	i \lambda_{R} & \Delta_{\text{sc}}  & v_{F} q_{-} & - \lambda_{I} & - i \lambda_{R} & 0 & 0 & 0 \\
\end{array}
\right)\ .
\end{align}

From the above dual Hamiltonian (Eq.~(\ref{SLG_BdGD})), we identify that the effective pairing term associated with nearest-neighbour lattice sites $(B_{\uparrow}^{\dagger}A_{\uparrow}^{\dagger})$ acquires a momentum dependence proportional to $q_{x}+iq_{y}$, indicating the emergence of chiral $p_{x}+ip_{y}$-type pairing in the spin-up sector. An analogous pairing structure also appears for the spin-down sector, implying that both spin species 
inherit the same effective pairing symmetry under the dual transformation.

Since our model Hamiltonian for the effective tBLG-WSe$_2$ is constructed from stacked graphene layers in the presence of SOCs, we argue that the emergence of such effective $p_{x}+ip_{y}$-type superconducting pairing within individual graphene layers naturally extends to the entire heterostructure via tunneling. 
In other words, the effective pairing symmetry generated at the single-layer level provides insight into the superconducting character of the full tBLG-WSe$_2$ system.

Furthermore, it is important to note that our BdG construction considers only the valley-$K$ sector in the particle block. This approximation is motivated by the fact that tBLG exhibits extremely weak intervalley scattering, allowing the two valleys to be treated approximately as independent degrees of freedom~\cite{MacDonald-tBLG,Koshino-tBLG}. Consequently, the graphene layers constituting the tBLG can also be analyzed within a single-valley framework, justifying our focus on the valley-$K$ description.

\section{Definitions: Bloch states and Chern number}\label{Sec:IV}
\begin{figure*}[t]
	\centering
	\subfigure{\includegraphics[width=0.75\textwidth]{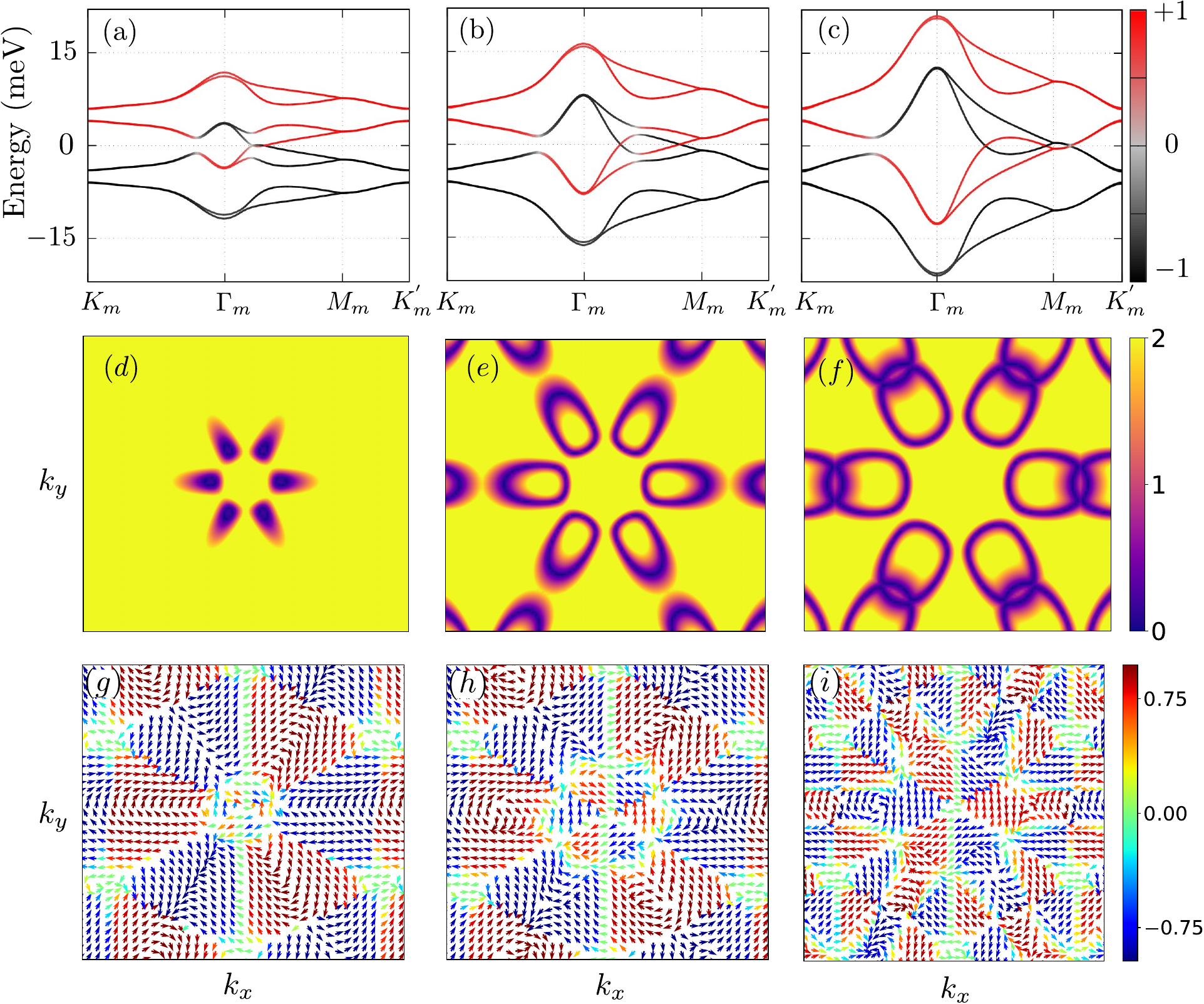}}
	\caption{The bulk band structure obtained from the BdG Hamiltonian of tBLG is depicted along the high-symmetry path in panels (a)-(c) for twist angles $\theta = 1.16^{\circ}, 1.20^{\circ},$ and $1.24^{\circ}$, respectively, choosing fixed Ising and Rashba SOC strengths $\lambda_{I} = 5$ meV and $\lambda_{R} = 10$ meV. On the other hand, panels (d)-(f) display the corresponding density plots of the band gap (in meV) over the reciprocal space for the same twist angles, while maintaining all other model parameters unchanged. The spin textures of Bloch states over the mBZ for these parameter sets are presented in panels (g)-(i). In all calculations, the superconducting order parameter is fixed at $\Delta_{\mathrm{sc}} = 1$ meV.
	}
	\label{tBLG_gapless_phase}
\end{figure*}

In this section, we introduce the entities used in our analysis, namely the Bloch states of the moir\'e system and the Chern number.
The real-space Bloch eigenstates corresponding to the low-energy continuum Hamiltonian can be written as
\begin{equation}
	\Psi^{s}_{n\mathbf k}(\mathbf r)
	= \sum_{\mathbf G^{m}} C^{s}_{n\mathbf k}(\mathbf G^{m}) \; e^{\,i\,(\mathbf k+ \mathbf G^{m})\!\cdot\mathbf r}\ ,
	\label{Eq:Bloch_state}
\end{equation}
where $\mathbf k$ denotes the Bloch wave vector, $\mathbf G^{m}$ represents the reciprocal lattice vectors of the moir\'e superlattice, $n$ is the band index, and $s$ labels the combined sublattice and layer degrees of freedom.

For multiband systems, especially in the presence of band crossings or nearly degenerate bands, the non-Abelian formalism of the Berry connection and Berry curvature provides the appropriate framework for evaluating the total Chern number associated with a group of bands (for \eg from the $\mu^{\rm{th}}$
to the $\nu^{\rm{th}}$ band). In this case, we define~\cite{non-Abelian_Berry_curvature_2019}
\begin{equation}
	C_{n}
	= \frac{1}{2\pi} \int_{\rm mBZ} d^{2}\mathbf k\;\mathrm{Tr}\!\bigl(\mathbf F_{n,\mathbf k}\!\cdot\hat{\mathbf z}\bigr)\ ,
	\label{Eq:Fukui-chern-1}
\end{equation}
where, the Berry curvature matrix $\mathbf F_{n\mathbf k}^{\,\mu \nu}$ is given by~\cite{non-Abelian_Berry_curvature_2019,Fukui-chern_no}
\begin{align}
	\mathbf F_{n\mathbf k}^{\,\mu \nu}
	&= \nabla_{\mathbf k}\!\times\! \mathbf A_{n\mathbf k}^{\,\mu \nu}
	+ i\bigl[A_{n,k_x},A_{n,k_y}\bigr]^{\mu \nu}\ , \\
	\mathbf A_{n\mathbf k}^{\,\mu \nu}
	&= i\,\langle \psi_{\mu\mathbf k}\mid \nabla_{\mathbf k}\mid \psi_{\nu\mathbf k}\rangle\ .
	\label{Eq:Fukui-chern-3}
\end{align}
Here, $|\psi_{\nu\mathbf k}\rangle$ denotes the eigenstate of the BdG Hamiltonian with band index $\nu$. The above non‐Abelian formulation is valid when the highest band in the set (\eg the $\nu^{\rm{th}}$ band) is separated by a direct gap from the $(\nu+1)^{\rm{th}}$ band. In our work, the Chern numbers presented in all topological phase diagrams of the main text, are obtained by summing over all occupied valence bands.
	
For our numerical computation, we use the gauge-invariant discretization scheme introduced by Fukui~\textit{et al.}~\cite{Fukui-chern_no} for evaluating the Chern number. Throughout this work, the Chern numbers appearing in the topological phase diagrams (see the main text) are computed by considering all occupied valence bands lying below the charge neutrality point of the BdG Hamiltonian.
	
\section{Nodal Loop Superconductivity in tBLG}\label{Sec:V}
In this section, we closely examine the gapless phase obtained in Fig.~2(d) of the main text. To better understand this phase, we calculate the band dispersion, the band-gap distribution over the mBZ, and the corresponding spin texture of the eigenstates on the mBZ.

In Figs.~\ref{tBLG_gapless_phase}(a)-(c), we present the band structures along the high-symmetry path $K_{m} - \Gamma_{m} - M_{m} - K_{m}^{'}$ of the mBZ for twist angles $\theta = 1.16^{\circ}, 1.20^{\circ}$, and $1.24^{\circ}$, respectively. At $\theta = 1.16^{\circ}$, the lowest two bands near the charge-neutral point begin to touch each other along the $\Gamma_{m} - M_{m}$ direction. With increasing the twist angle, the band crossing becomes more prominent, leading to clearly visible zero-energy crossings at two distinct momentum points for $\theta = 1.20^{\circ}$ and $1.24^{\circ}$. These crossings manifest as nodal loops 
in the momentum-resolved band-gap distribution, defined as the energy difference between the lowest conduction and highest valence bands.

On the other hand, Figs.~\ref{tBLG_gapless_phase}(d)-(f) exhibit the evolution of these nodal loops over the mBZ for the same sequence of twist angles. When $\theta = 1.16^{\circ}$, vanishingly small isolated low-gap pockets appear around six symmetry-related momentum points. At $\theta = 1.20^{\circ}$, these pockets expand into elongated closed contours, which eventually merge into interconnected nodal structures at $\theta = 1.24^{\circ}$. In all our calculations, the 
Ising and Rashba SOC strengths are fixed at $\lambda_{I} = 5$ meV and $\lambda_{R} = 10$ meV, respectively, while the superconducting order parameter is set 
to $\Delta_{\mathrm{sc}} = 1$ meV.

Since the emerging superconducting phase becomes nodal, the Chern number is no longer well defined to characterize such phase. To gain insight into the topological character of this gapless phase, we therefore examine the spin texture over the mBZ and display its behavior in Figs.~\ref{tBLG_gapless_phase}(g)-(i) 
for the same twist angles as mentioned before. 
Here, the arrow directions are determined by the expectation values of the in-plane spin components 
$(\langle s_x \rangle, \langle s_y \rangle)$, while the color scale represents the out-of-plane component $\langle s_z \rangle$. The emergence of gapless nodal structure is accompanied by a significant reconstruction of the spin texture, indicating an SOC-driven reorganization of the pseudospin winding. 
These results suggest a transition from a fully gapped topological superconducting phase to a nodal superconducting state.  

\section{Role of Intrinsic SOC in Stabilizing Topological Superconductivity}\label{Sec:VI}
\begin{figure}[h]
	\centering
	\subfigure{\includegraphics[width=0.48\textwidth]{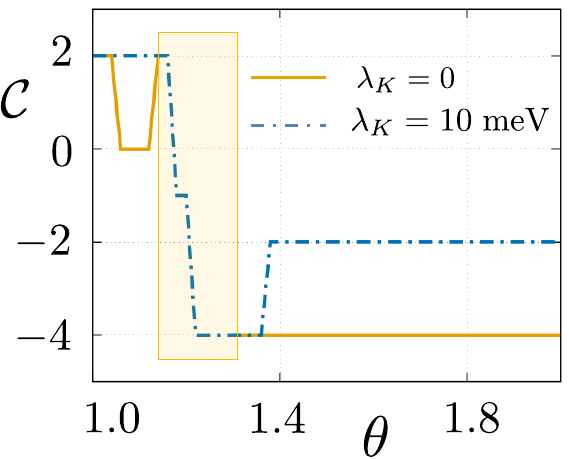}}
	\caption{Chern number ($\mathcal{C}$) is depicted as a function of the twist angle ($\theta$), in presence 
	of intrinsic SOC strengths $\lambda_{K}=0$ (yellow line) and $\lambda_{K}=10$ meV (blue line). The Ising and Rashba SOC strengths are considered as $\lambda_{I}=5$ meV and $\lambda_{R}=10$ meV, respectively, while the superconducting order parameter is set to $\Delta_{\mathrm{sc}}=1$ meV.
	}
	\label{effect of ISOC}
\end{figure}

To elucidate the role of intrinsic SOC in shaping the topological superconducting phases, we consider two scenarios: in the absence and presence of intrinsic (Kane--Mele type) SOC~\cite{Kane-mele_QSH1,Kane-mele_QSH2}, characterized by $\lambda_K=0$ and $10$ meV, respectively. In Fig.~\ref{effect of ISOC}, we show the Chern number ($\mathcal{C}$) as a function of the twist angle ($\theta$) in presence of finite $s$-wave superconducting pairing, Ising SOC, and Rashba SOC strengths. The yellow and blue curves correspond to $\lambda_K=0$ and $10$ meV, respectively.

In the absence of intrinsic SOC, the system exhibits an extended gapless region, highlighted by the transparent yellow box, where the bulk gap closes and the Chern number becomes ill-defined. Consequently, the topological characterization of the superconducting phase cannot be achieved within this parameter range. In contrast, upon incorporating finite intrinsic SOC while keeping all other model parameters unchanged, the gapless region disappears and several additional gapped topological phases with integer Chern number emerge. This demonstrates that intrinsic SOC plays a crucial role in stabilizing gapped topological superconducting phases and enriching the topological phase diagram.
Note that, the suppression of the gapless phase requires a critical strength of the intrinsic SOC which 
needs to exceed that of the Ising SOC. When $\lambda_K > \lambda_I$, the bulk gap remains finite over a broader range of twist angles, thus enabling appearance of well-defined topological phases characterized by distinct Chern numbers.

\end{onecolumngrid}	
\end{document}